  \documentclass[aps,prl,twocolumn,superscriptaddress,floatfix,amsmath]{revtex4-1}
 \usepackage{graphicx}
 \usepackage{amsmath}
 \usepackage{amssymb}
 \usepackage{textcomp} 
\usepackage[dvips]{color}
 \usepackage[normalem]{ulem} 
 \usepackage{hyperref}
\definecolor{g-blue}{rgb}{0.83,0.95,1}
\definecolor{g-yellow}{rgb}{1,1,0.7}
\definecolor{g-green}{rgb}{0.9,1,0.9}
\definecolor{green}{rgb}{0,0.6,0}
\definecolor{cyan}{rgb}{0,0.7,0.7}
\definecolor{black}{rgb}{0,0,0}
\definecolor{grey}{rgb}{0.4 ,0.4 ,0.4 }

\usepackage{bm}
\usepackage{amssymb}
\usepackage{amsfonts}
\usepackage{amsmath}
 \usepackage{graphicx}
  \usepackage{amsmath,bm,epsfig}
\def \ed {\end{document}}
\def\Fbox#1{\vskip1ex\hbox to 8.5cm{\hfil\fboxsep0.3cm\fbox{%
  \parbox{8.0cm}{#1}}\hfil}\vskip1ex\noindent}  

\newcommand{\Ref}[1]{Ref.\,\cite{#1}}

\def\be{\begin{equation}}\def\ee{\end{equation}}
\def\bea{\begin{eqnarray}}\def\eea{\end{eqnarray}}
\def\bse{\begin{subequations}}\def\ese{\end{subequations}}
\newcommand{\BE}[1]{\begin{equation}\label{#1}}
\newcommand{\BEA}[1]{\begin{eqnarray}\label{#1}}
\newcommand{\BSE}[1]{\begin{subequations}\label{#1}}

\let \= \equiv \let\*\cdot \let\~\widetilde \let\^\widehat \let\-\overline

  \def\1{\bm1} 

\def\<{\left\langle}    \def\>{\right\rangle}
\def\({\left(}          \def\){\right)}
 \def \[ {\left [} \def \] {\right ]}

\newcommand{\B}[1]{{\bm{#1}}}

\def\He4 {$^4$He~}
\begin{document}

\title{On supercurrents in Bose-Einstein magnon condensates in a YIG ferrimagnet}
\author{Dmytro~A.~Bozhko}
\affiliation{Fachbereich Physik and Landesforschungszentrum OPTIMAS, Technische Universit\"at Kaiserslautern, 67663 Kaiserslautern, Germany}
\affiliation{Graduate School Materials Science in Mainz, 67663 Kaiserslautern, Germany}

\author{Alexander~A.~Serga}
\affiliation{Fachbereich Physik and Landesforschungszentrum OPTIMAS, Technische Universit\"at Kaiserslautern, 67663 Kaiserslautern, Germany}

\author{Peter Clausen}
\affiliation{Fachbereich Physik and Landesforschungszentrum OPTIMAS, Technische Universit\"at Kaiserslautern, 67663 Kaiserslautern, Germany}

\author{Vitaliy~I.~Vasyuchka}
\affiliation{Fachbereich Physik and Landesforschungszentrum OPTIMAS, Technische Universit\"at Kaiserslautern, 67663 Kaiserslautern, Germany}

\author{Gennadii~A.~Melkov}
\affiliation{Faculty of Radiophysics, Electronics and Computer Systems, Taras Shevchenko National University of Kyiv, Kyiv 01601, Ukraine}

\author{Anna~Pomyalov}
\affiliation{Department of Chemical Physics, Weizmann Institute of Science, Rehovot 76100, Israel \looseness=-1}

\author{Victor~S.~L'vov}
\affiliation{Department of Chemical Physics, Weizmann Institute of Science, Rehovot 76100, Israel \looseness=-1}

\author{Burkard~Hillebrands}
\email{hilleb@physik.uni-kl.de}
\affiliation{Fachbereich Physik and Landesforschungszentrum OPTIMAS, Technische Universit\"at Kaiserslautern, 67663 Kaiserslautern, Germany}
\begin{abstract}
Recently  E. Sonin commented\,\cite{Sonin}  on our preprint ``Supercurrent in a room temperature Bose-Einstein magnon condensate'' \cite{NatPhysArx,NatPhys},
   arguing that our ``claim  of detection of spin supercurrent is premature and has not been sufficiently supported by presented experimental results and their theoretical interpretation."
We consider the appearance of this Comment as a sign of significant interest into the problem of supercurrents in  Bose-Einstein magnon condensates.
Here, we explicitly address E. Sonin's comments and show  that our interpretation of our experimental results as a  detection of a magnon supercurrent is fully supported not only by the experimental results themselves, but also by independent theoretical analysis  \cite{Pokr}.

\end{abstract}
\maketitle

Recently E. Sonin submitted a short Comment on our preprint ``{\it Supercurrent in a room temperature Bose-Einstein magnon condensate}"\cite{NatPhysArx,NatPhys}, to arXiv\,\cite{Sonin},  which we have   addressed  in a previous Note\,\cite{reply1}.
Here we discuss the four main statements of \Ref{Sonin} in more details.
\smallskip

 {\it Statement I. ``Observation of spin current in Yttrium-Iron-Garnet (YIG) is hardly possible''}.

 This statement is based on two criteria for the existence of a steady supercurrent, formulated in \Ref{2}    in terms directly applicable for easy-plane (anti)ferromagnets, where the minimum of the magnon frequency $\omega(\B k)$ corresponds to $k=0$ and BEC is nothing else but the homogeneous precession of the magnetization.  The situation in tangentially magnetized YIG films is quite different: the dispersion relation  $\omega(\B k)$ has two minima at non-zero wavevector values $\pm \B k_0$ (with $k_0\simeq 5\cdot 10^4\,$cm$^{-1}$) and the BEC spate is a magnon state with a wavelength of $\lambda \simeq 10^{-4}\,$ cm,  which at least ten times smaller than the hot-spot radius.  In this case we have no problem with the anisotropy in the complex plane of the BEC phase (see e.g. \Ref{Pokr}) and the magnon phase variation across  the hot spot exceeds $2\pi$ at least ten times.

\smallskip

{\it Statement II. ``The authors did not consider a more common scenario of spin diffusion, which could \underline{probably} successfully compete with their present interpretation. It is worth noting that spin diffusion would be more effective in the condensed state than in a non-condensed gas of magnons''}.

 We  present below   evidence that spin diffusion cannot be  observed  in our experiments and theoretical arguments that the spin diffusion is much smaller than the spin supercurrent.
\paragraph{Experimental arguments.} We explained in page 4 of our preprint, that ``the laser heating process decreases locally the saturation magnetization via thermal excitation of high-energy magnons with terahertz frequencies, but practically does not increase the local population of low-energy magnons. As a result, the number of thermal magnons in the low-energy spectral region remains negligibly small in comparison with the number of magnons originating from the parametric pumping process and cannot visibly affect the BEC dynamics''.
Moreover, we observed a critical value of the bottom-magnon populations $N_\mathrm{cr}$
(see horizontal dashed line in Fig.~5) that separates two regions with different spin dynamics; in the upper region the only reasonable explanation for the observed spin dynamics is the existance of a supercurrent of magnons within the BEC, because a diffusive spin current is not a threshold effect, and, thus, cannot lead to  the appearance of $N_\mathrm{cr}$. Thus, we  conclude: \textit{The above described experimental facts exclude the possibility of a significant contribution of the spin diffusion current to the observed spin dynamics}. \looseness=-1

\paragraph{Theoretical arguments.}   To support the above  conclusion theoretically, we could have chosen to present an analytical estimation of the relative roles of the magnon supercurrent and a normal diffusion current. Such an estimate has been discussed in a recent PRL paper by C.\,Sun, T. \,Nattermann and V.\,Pokrovsky \cite{Pokr} with the following result: ``Thus, we expect that, in realistic circumstances, the spin superfluid current (in YIG films) equal to $10^{22}\div 10^{23}\,$ cm$^{-2}$s$^{-1}$ \emph{is larger than the normal current by $(3\div 5)$ decimal orders}''.  Due to experimental limitations regarding the apparatus calibration, we cannot give the absolute value  of the spin supercurrent, but we believe that for the purpose of our discussion the relative estimate, obtained in \Ref{Pokr}, is sufficient.

\smallskip

{\it Statement III. "The authors investigated not stationary, but time dependent spin currents."}

 Indeed, our experiments are dealing with   time-dependent decay processes with a characteristic decay of time about $(3\div 4)\cdot 10^{-7}\,$s. This time is at least 10 times longer than the characteristic time of the four-magnon interaction, responsible for the creation of the magnon BEC state. Therefore we believe that our BEC can be considered as being quasi-stationary,  at least in  first-order approximation. However we agree with E. Sonin that an estimate of the effect of the non-stationarity is important in a consistent theory of the magnon BEC phenomenon.  We hope to address this problem during ongoing research.

{\it Statement IV. "The agreement \emph{(in our Ref.\,\cite{NatPhysArx,NatPhys})} looks very good indeed, probably due to using a fitting parameter."}

We believe that such a good agreement stems from the fact, that our model reflects well the basic physics of the observed phenomena. If there exists an alternative quantitative theoretical interpretation of the experimental results, we would be happy to study it to compare, which model describes our experimental observations better.

 {\bf Conclusion.}
 We consider E. Sonin's Comment\,\cite{Sonin} as a sign of a wide interest to the subject  and thank him for the constructive discussions.
 	We believe that in this Note we clarified   the differences in the physical situations discussed in \Ref{Sonin, NatPhysArx, reply1}.  We are confident that our experimental results confirm the existence of a magnon supercurrent in a room-temperature magnon Bose-Einstein condensate.

\end{document}